\documentclass[aps,prb,preprint,superscriptaddress,showpacs]{revtex4}
\usepackage{graphicx}
\usepackage{mathrsfs}

\begin{document}




\title{Dissipative and conservative nonlinearity in carbon nanotube and graphene mechanical resonators}

\author{J. Moser}
\affiliation{Institut Catal\'{a} de Nanotecnologia, Bellaterra
08193, Spain}
\author{A. Eichler}
\affiliation{Institut Catal\'{a} de Nanotecnologia, Bellaterra
08193, Spain}

\author{B. Lassagne}
\affiliation{Institut Catal\'{a} de Nanotecnologia, Bellaterra
08193, Spain}

\author{J. Chaste}
\affiliation{Institut Catal\'{a} de Nanotecnologia, Bellaterra
08193, Spain}

\author{Y. Tarakanov}
\affiliation{Department of Applied Physics, Chalmers University of
Technology, G\"{o}teborg 41296, Sweden}

\author{J. Kinaret}
\affiliation{Department of Applied Physics, Chalmers University of
Technology, G\"{o}teborg 41296, Sweden}

\author{I. Wilson-Rae}
\affiliation{Technische Universit\"{a}t M\"{u}nchen, Garching 85748,
Germany}
\author{A. Bachtold}
\affiliation{Institut Catal\'{a} de Nanotecnologia, Bellaterra
08193, Spain}



\maketitle

\section{Introduction}

\begin{figure}[b]
\begin{center}
\includegraphics{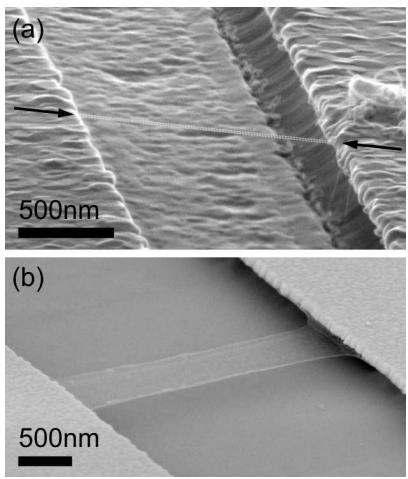}
\end{center}
\caption{(a) Scanning electron microscopy image of a nanotube
resonator. The nanotube was grown by chemical vapor deposition over
a prefabricated trench between two W/Pt contacts. The nanotube is
marked by the arrows and the white dotted lines. (b) Scanning
electron microscopy image of a suspended single-layer graphene sheet
with Au contacts. (Adapted with permission from
Ref.~\cite{naturenano}.)} \label{moser_fig1}
\end{figure}

Graphene and carbon nanotubes are excellent materials for
nanoelectromechanical systems (NEMS). Their large stiffness and low
density allow to fabricate high frequency mechanical resonators
sensitive to minute variations of mass, force, and charge
\cite{poncharal,reulet,purcell,babic,sazanova,witkamp,daniel,benjamin_nanolett,chiu,jensen,huttel,benjamin_science,steele,gouttenoire,wang,wu,alex_nanolett,bunch,daniel_nanolett,robinson2008,chen,singh,arend2010,barton2011,naturenano}.
In addition to their outstanding mechanical properties, these
materials owe part of their uniqueness to their simplest feature:
They constitute the ultimate size limit for one- and two-dimensional
(1D and 2D) NEMSs. Indeed, single-walled nanotubes are ultra-narrow
wires whose diameter can be as small as 1~nm; graphene, being a
sheet of carbon atoms arranged in a honeycomb lattice, is the
thinnest membrane imaginable. Figure~\ref{moser_fig1} shows
mechanical devices made from a carbon nanotube and a graphene sheet.

\begin{figure}[t]
\begin{center}
\includegraphics{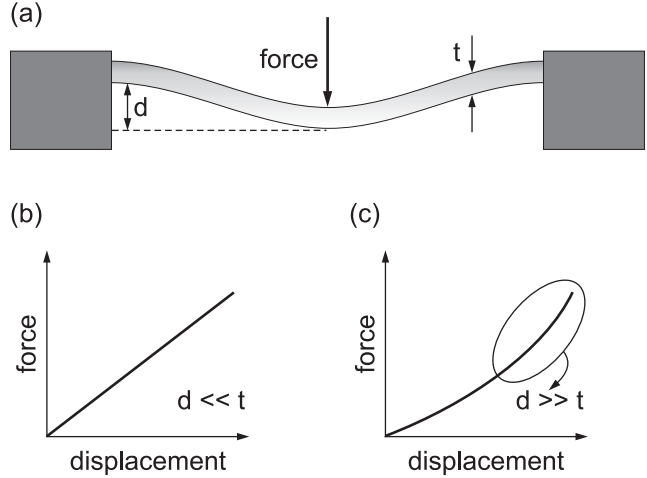}
\end{center}
\caption{(a) Schematic of a suspended beam clamped at both ends and
subject to a force. (b,c) Expected force-displacement
characteristics when the displacement $d$ is much lower than the
beam thickness $t$ (b) and when $d$ is much larger than $t$ (c).}
\label{moser_fig2}
\end{figure}

Owing to their reduced dimensionality, graphene and carbon nanotubes
display unusual mechanical behavior. We will discuss two examples in
the following. The first one concerns the force-displacement
response. In the well established case of the cantilever of an
atomic force microscope probe, its position changes linearly with
the force it is subjected to. Such a linear force-displacement
response is also found in doubly-clamped beams (see
Fig.~\ref{moser_fig2}a and b). According to continuum elasticity
theory, in the latter case this linearity is expected to hold only
when the displacement is much smaller than the thickness of the beam
\cite{Landau_Lifshitz}. Because the diameter of a nanotube and the
thickness of a graphene sheet are usually both small compared to any
displacement, force-displacement measurements in nanotubes and
graphene are expected to be highly nonlinear (such as in
Fig.~\ref{moser_fig2}c). This was indeed shown to be the case for
graphene \cite{lee}.

\begin{figure}[t]
\begin{center}
\includegraphics{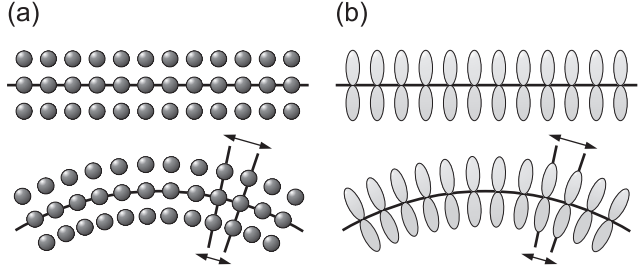}
\end{center}
\caption{(a) Arrangement of the atoms of a beam in a straight (top)
and bent (bottom) geometry. (b) Arrangement of the $p_{z}$
electronic orbitals of the carbon atoms of a graphene sheet in a
straight (top) and bent (bottom) geometry. The bending energy is
given by the energy cost for the electronic orbitals to rotate.}
\label{moser_fig3}
\end{figure}

Another striking example of an unusual mechanical property
originating from the reduced dimensionality of these carbon-based
resonators concerns the bending rigidity. This quantity
characterizes the resistance of an object to bending.
Figure~\ref{moser_fig3}a shows that the lower region of a bent beam
contracts while the upper region expands (when the beam bends
upwards). At the microscopic level, the separation between atoms is
larger (smaller) in the upper (lower) region. This is energetically
not favorable, so the beam tends to return to its initial, straight
configuration. In this respect, the case of graphene is intriguing
(Fig.~\ref{moser_fig3}b): because it is only one atom thick, the
bending rigidity results solely from the energy cost of changing the
angle between the $p_z$ orbitals of carbon atoms \cite{atalaya}.
Thus the microscopic origin of the bending rigidity of graphene
differs from the one of standard materials. This book chapter
focuses on graphene and nanotube NEMS resonators and on the peculiar
mechanical properties that their reduced dimensionality entails.

From a practical point of view, the low dimension and the
corresponding ultra-low mass of nanotubes and graphene present a
great advantage in such experiments as inertial mass sensing and the
exploration of the quantum regime of a macroscopic mechanical degree
of freedom. The working principle of mass sensing is simple.
Electromechanical resonators can be described as harmonic
oscillators with an effective mass $m$ (close to the real mass of
the resonator volume), a spring constant $k$, and a mechanical
resonant frequency $f_{0}=1/2\pi\cdot\sqrt{k/m}$. Mass sensing
consists in monitoring the shift in $f_0$ induced by the adsorption
of atomic species onto the resonator. The reason for the high mass
sensitivity of nanotube resonators is that the mass of a nanotube is
ultra-low, so even a tiny amount of atoms deposited on the nanotube
makes up a significant fraction of the total mass. Whence, it has
been possible to achieve a mass sensitivity of about $1~{\rm{zg}} =
10^{-21}$~g with a nanotube resonator
\cite{benjamin_nanolett,chiu,jensen}, which surpasses the
sensitivity achieved using resonators based on other materials
\cite{yang}. Note that the frequency shift in inertial mass sensing
depends not only on the adsorbed mass but also on the position of
the adsorbate. By exploiting the nonlinear mechanics of nanoscale
resonators, the position can in principle be determined by a
measurement at the fundamental resonant frequency by using a
multi-frequency excitation scheme \cite{kinaret}.

Their low mass makes carbon nanotubes and graphene very promising
for the study of their motion in the quantum regime. This in
practice requires preparing the resonator mode (oscillator) close to
its quantum ground state. What makes such experiments a
technological feat is that the amplitude of the zero-point motion is
typically very low ($x_{zp}=\sqrt{\hbar/4\pi f_{0}m}$), and
therefore difficult to detect. In this context, graphene and
nanotubes offer the immediate advantage of a very low mass that
renders the zero-point motion larger than in heavier resonators. A
ballpark figure for $x_{zp}$ is typically $1-10$~fm for resonators
microfabricated from semiconducting and metallic materials and
operated in the $1-10$~MHz range
\cite{lenhert_prl,rocheleau_nature}. By contrast, $x_{zp}$ is
expected to be $1-10$~pm for graphene resonators with similar
resonant frequencies. Not only does a large zero-point motion make
detecting the motion in the quantum limit easier, it also allows for
an enhanced coupling to other degrees of freedom, such as the
photons of a superconducting resonator
\cite{lenhert_prl,rocheleau_nature} and the qubit in a Cooper-pair
box \cite{lahaye_nature}. This coupling to nonlinear elements such
as two level systems is needed to observe quantum dynamics;
interestingly, however, graphene resonators are nonlinear at very
small oscillation amplitudes, opening the possibility to study
quantum dynamics without external components \cite{voje}. We note
that the quantum dynamics of nonlinear vibrations was investigated
in the past by means of spectroscopy measurements of the vibrations
of impurities in solids \cite{elliott}.

\section{Detection of the mechanical vibrations}

Even though carbon nanotubes and graphene are outstanding materials
for NEMS resonators, what first hindered progress in this research
field is the difficulty to detect their mechanical vibrations.
Nanotubes and graphene are so small that it is challenging to apply
the detection schemes used for larger mechanical resonators
\cite{Roukes_Physics_World}. The first measurements were carried out
by transmission and scanning electron microscopy
\cite{poncharal,babic} as well as field emission \cite{purcell}, but
these detection schemes are suitable neither for sensing nor for
experiments in the quantum limit. Over the last few years, efforts
have been made to develop new detection schemes based e.g. on
optical interferometry \cite{bunch}, atomic force microscopy
\cite{daniel,daniel_nanolett}, electrical \cite{huttel}, and
capacitive \cite{eric,hone} readouts. A technique that has become
very popular is the mixing technique. It essentially consists of
measuring the electrical current flowing through the graphene sheet
or the nanotube. This detection scheme is very practical since it
can be implemented in various experimental setups, such as cryostats
and ultra-high vacuum chambers. The method was first used in
experiments on microfabricated resonators \cite{cleland_nature} and
was later on adapted to nanotube resonators by the McEuen group at
Cornell \cite{sazanova}. The mixing technique was recently improved
by the group in Lyon \cite{gouttenoire} and we will discuss this
improved version.

\begin{figure}[t]
\begin{center}
\includegraphics{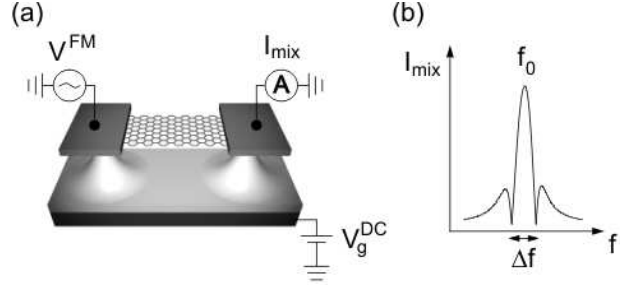}
\end{center}
\caption{(a) Schematic of the device layout and the
actuation/detection setup. (b) Schematic of the frequency response
of the mixing current. The separation between the two minima
corresponds to the resonance width $\Delta f$, equal to $f_{0}/Q$.
(Adapted with permission from Ref.~\cite{naturenano}.)}
\label{moser_fig4}
\end{figure}

The resonator is actuated electrostatically by applying an
oscillating voltage $V^{FM}$ with amplitude $V^{AC}$ to the source
electrode, which creates an alternating potential between the
resonator and the gate electrode (Fig.~\ref{moser_fig4}a). Central
to the technique is the fact that this driving voltage is frequency
modulated (FM) to take the form $V^{FM}=V^{AC}\cos[2\pi
ft+f_{\Delta}/f_{L}\cdot\sin(2\pi f_{L}t)]$, where $f$ is the
carrier frequency and $f_{\Delta}$ is called frequency deviation.
The motion is detected by measuring the current at frequency $f_{L}$
at the drain electrode, $I_{\rm{mix}}\propto |\partial
{\rm{Re}}[x_{0}]/\partial f|$, where ${\rm{Re}}[x_{0}]$ is the real
part of the frequency response function of the displacement $x_{0}$.
The expression $|\partial{\rm{Re}}[x_{0}]/\partial f|$ can be
understood as follows. It contains an absolute value because
$I_{\rm{mix}}$ is measured with a lock-in amplifier and the output
signal is the module of the current. The derivative with respect to
$f$ is a consequence of the modulation of the applied voltage:
$V^{FM}$ can be seen as an oscillating voltage whose frequency is
modulated as $f_{i}=f+f_{\Delta}\cos (f_{L}t)$ (valid over a
timescale shorter than $Q/f_{0}$, where $Q$ is the quality factor
and $f_{0}$ the resonant frequency of the mechanical resonator).
Finally, the origin of the real part in $|\partial
{\rm{Re}}[x_{0}]/\partial f|$ is less transparent; as shown in
\cite{gouttenoire}, it comes from the Taylor expansion of
$I_{\rm{mix}}$ at small motional amplitude $\delta x(t)$ and small
drive voltage $\delta V(t)$, whose only component at frequency
$f_{L}$ is contained in the term $\partial ^{2}I_{\rm{mix}}/\partial
x
\partial V \cdot \delta x \delta V$.

One advantage of this measurement technique is the frequency
conversion. The frequency of the measured signal ($f_{L}$) is about
1~kHz whereas the motion of the resonator can have much higher
frequencies (typically $10-1000$~MHz). This frequency conversion is
essential since the impedance of nanotube and graphene resonators is
much larger than 50~Ohm, which makes it difficult to measure small
high-frequency currents through the resonator. A further advantage
of this detection scheme is that the mixing current as a function of
$f$ has a characteristic line-shape (Fig.~\ref{moser_fig4}b), which
allows us to extract the mechanical quality factor $Q$ in a simple
manner: the resonance peak at frequency $f_0$ is flanked by two
minima whose separation is the resonance width $\Delta f = f_{0}/Q$
for a linear harmonic oscillator. Later on we will see that the
distance between these two minima is also a natural measure of $Q$
in a nonlinear oscillator.

Choosing the right value for the frequency deviation ($f_{\Delta}$)
of the FM technique is crucial for a reliable measurement. Namely,
one has to ensure that $f_{\Delta}$ is sufficiently small compared
to the width of the mechanical resonance $\Delta f$ (typically,
$f_{\Delta}<\Delta f/4$). Otherwise, the measured resonance broadens
\cite{gouttenoire,naturenano}. (In practice, we measure the
dependence of $\Delta f$ on $f_{\Delta}$; the real resonance width
is obtained at low $f_{\Delta}$ where $\Delta f$ saturates).

\section{Variation of the quality factor with the amplitude of the
motion}\label{moser_section3}

Large mechanical resonators, such as those microfabricated from
metallic or semiconducting materials, are usually well described as
simple harmonic oscillators. The equation of motion is then given by
the well-known expression:
\begin{equation}
md^{2}x/dt^{2}=-kx-\gamma dx/dt + F_{\rm{drive}}\cos(2\pi ft)
\label{moser_eq1}
\end{equation}
which interrelates the position $x$, the velocity $dx/dt$, and the
acceleration $d^{2}x/dt^{2}$ of the oscillator (with effective mass
$m$, spring constant $k$, and damping rate $\gamma$). Here, we show
that the motion of graphene/nanotube resonators is different from
that of larger mechanical resonators \cite{naturenano}. The most
striking difference is that the damping depends on the amplitude of
the motion, in sharp contrast to the behavior of simple harmonic
oscillators.

A surprising experimental fact in nanotube and graphene resonators
is that the mechanical resonance lineshape often broadens as the
driving force is increased
($F_{\rm{drive}}=C^{\prime}V_{g}^{DC}V^{AC}$ where $C^{\prime}$ is
the derivative of the gate-resonator capacitance with respect to
$x$). In other words, the quality factor depends on the amplitude of
the motion. This is a novel phenomenon: indeed, in larger resonators
the quality factor is independent of the motional amplitude ($Q=2\pi
f_{0}m/\gamma$).

\begin{figure}[b]
\begin{center}
\includegraphics{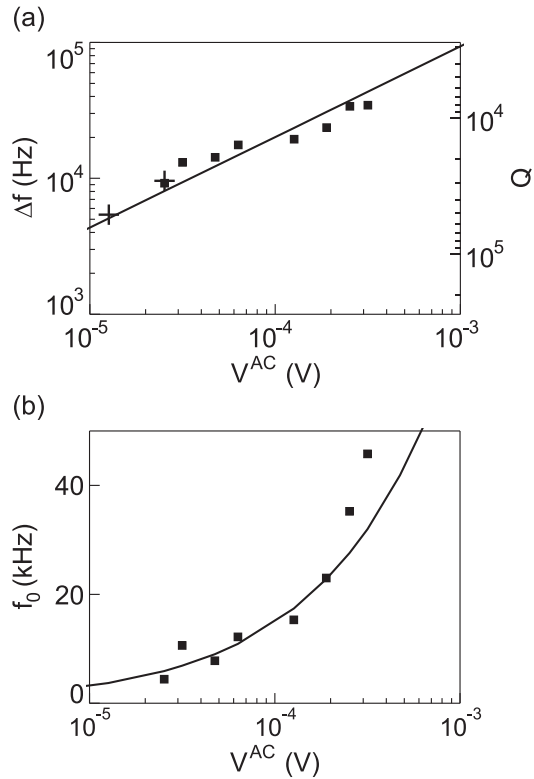}
\end{center}
\caption{Measurements of the mechanical properties of a resonator
made from a nanotube under tension. The length of the nanotube is
840~nm and the radius is 2~nm. (a) Resonance width as a function of
drive $V^{AC}$. Squares correspond to 5~K and crosses to 400~mK. The
line corresponds to equation~(\ref{moser_eq5}) with
$\eta=10^{4}$~kg$\cdot\rm{m}^{-2}\cdot\rm{s}^{-1}$ ($\gamma=0$). The
corresponding quality factor $Q$ is shown on the right-hand side
scale. (b) Resonance shift as a function of $V^{AC}$. The line is a
solution to equation~(\ref{moser_eq3}) with
$\alpha=6\cdot10^{12}$~kg$\cdot\rm{m}^{-2}\cdot\rm{s}^{-2}$.
(Adapted with permission from Ref.~\cite{naturenano}.)}
\label{moser_fig5}
\end{figure}

Figure~\ref{moser_fig5}a shows an example of the variation of the
resonance width (and of the quality factor on the right axis) as a
function of driving force (which scales linearly with $V^{AC}$) in a
logarithmic plot. The variation of the quality factor is significant
and can reach a factor of 50 for some devices. This behavior is
robust as it can be observed both in nanotube and graphene
resonators, both at low temperature (down to 90~mK) and at room
temperature, and using different detection methods (namely, the FM
method described in the previous section, the so-called two source
technique \cite{sazanova}, and optical interferometry \cite{bunch}).

The dependence of the mechanical quality factor on the amplitude of
the driving force came as a surprise to us. A single mention of this
phenomenon appears in \cite{bunch}, but its origin is not discussed.
One possible reason why this effect remained unnoticed before is
that the variation of the quality factor becomes clearly visible
only over a rather large span of driving force amplitudes, while the
detection of the motion at low drive is challenging. In these
measurements, care is taken to avoid driving $V^{AC}$ above
$k_{\rm{B}}T/e$ in order to prevent electronic nonlinear effects or
local heating (here $k_{\rm{B}}$ is the Boltzmann constant, $T$ the
temperature, and $e$ the electron charge). (On a technical note, the
variation of the quality factor does not stem from the coupling
between electrons and mechanical vibrations
\cite{benjamin_science,steele} because the effect is not associated
to Coulomb blockade and $V^{AC}$ is kept below $k_{\rm{B}}T/e$, see
Ref.~\cite{naturenano} for a more detailed discussion.)

In order to understand the variation of the quality factor with the
drive, it is useful to first discuss another behavior. Increasing
the driving force also shifts the resonant frequency to higher (or
sometimes lower) values (Fig.~\ref{moser_fig5}b). This behavior is
usually associated to the so-called Duffing force
$F_{\rm{Duffing}}=-\alpha x^{3}$ (Ref.~\cite{cross_lifshitz}). The
latter, together with the spring force, can be expressed as
\begin{equation}
-k\left(1+\frac{\alpha}{k}x^{2}\right)x \label{moser_eq2}
\end{equation}
which indicates that the Duffing term contributes to the restoring
force: it makes the resonator stiffer (for $\alpha>0$) and increases
the resonant frequency. However, the Duffing force alone cannot
explain the variation of the quality factor with drive, since it
does not affect the resonance width (in the absence of any
bistability), provided that the amplitude of thermal fluctuations is
smaller than that of the driven motion \cite{dykman2}. This is
readily verified by adding the Duffing force to the right hand side
of equation~(\ref{moser_eq1}) and solving for the steady state. The
resonance width, given by the separation between the two minima in
the mixing current as a function of $f$ in the FM technique, is
indeed independent of the driving force. In addition to the Duffing
nonlinearity $-\alpha x^{3}$, the other relevant higher-order term
in the Newton equation for a weakly anharmonic oscillator is the
nonlinear damping term $-\eta x^2 dx/dt$
(Ref.~\cite{dykman,cross_lifshitz,stav}). We thus get
\begin{equation}
md^{2}x/dt^{2}=-kx-\gamma dx/dt-\alpha x^{3}-\eta
x^{2}dx/dt+F_{\rm{drive}}\cos(2\pi ft) \label{moser_eq3}
\end{equation} A derivation of equation~(\ref{moser_eq3}) based on
the Caldeira-Leggett model can be found in \cite{stav}, in which the
nonlinear damping force emerges from a nonlinear coupling between
the mechanical resonator and a thermal bath of harmonic degrees of
freedom (the nonlinearity in the coupling results from the
anharmonicity of the potential well in which the resonator is
confined). Other works showed that additional terms of second and
third order ($x^{2}$, $xdx/dt$, $(dx/dt)^{2}$, $x(dx/dt)^{2}$,
$(dx/dt)^{3}$) lead in the rotating frame to a renormalization of
$\alpha$ and $\eta$ (Ref.~\cite{dykman2,dykman,cross_lifshitz}).

The term $-\eta x^{2}dx/dt$ is a damping force since it scales
linearly with the velocity. Together with $-\gamma dx/dt$, it can be
expressed as
\begin{equation}
-\gamma\left(1+\frac{\eta}{\gamma}x^{2}\right)dx/dt
\label{moser_eq4}
\end{equation}
The force $-\eta x^{2}dx/dt$ accounts for a dissipation mechanism
that becomes important at large motional amplitude. When $-\gamma
dx/dt$ dominates over $-\eta x^{2}dx/dt$, which is the case for
larger mechanical resonators, the resonance width is independent of
the driving force and is given by $\Delta f=\gamma/2\pi m$. In the
opposite limit \cite{naturenano}, when the term $-\gamma dx/dt$ can
be neglected,
\begin{equation}
\Delta f=0.032m^{-1}\eta^{1/3}f_{0}^{-2/3}F_{\rm{drive}}^{2/3}
\label{moser_eq5}
\end{equation}
so that $\Delta f\propto (V^{AC})^{2/3}$. At first sight, the simple
relation $Q=f_{0}/\Delta f$ for a linear harmonic oscillator is
expected to break down in the presence of nonlinearities. However,
we show in \cite{naturenano} that an analogous expression
$Q=1.09f_{0}/\Delta f$ is recovered in the limit of strong nonlinear
damping.

The relation $\Delta f\propto (V^{AC})^{2/3}$ captures well the
experimental data:  the line in the double logarithmic plot of
Fig.~\ref{moser_fig5}a corresponds to a power law with an exponent
2/3. This fit allows to extract $\eta$. The agreement between
experiment and theory is a strong indication that damping is here
described by the nonlinear force $-\eta x^{2}dx/dt$, instead of the
linear force $-\gamma dx/dt$.

In the following we discuss two additional experimental facts which
further support the dominance of the nonlinear damping force in
nanotube/graphene resonators. These are revealed by studying the
hysteretic behavior of the mechanical resonance on one hand and
parametric excitation on the other.

\section{Hysteresis}

\begin{figure}[t]
\begin{center}
\includegraphics{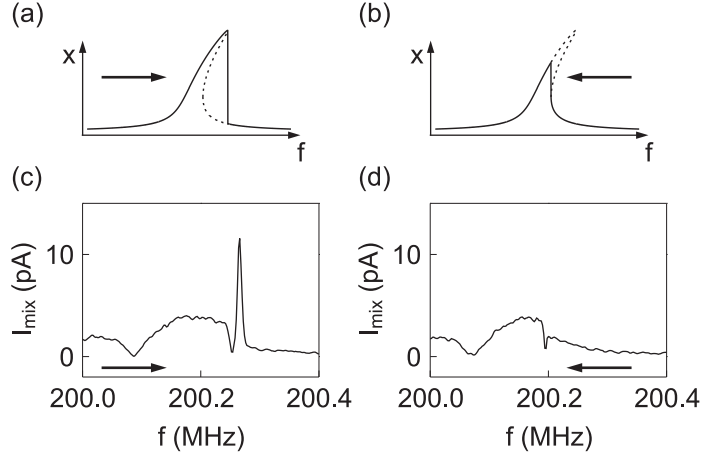}
\end{center}
\caption{Mechanical properties of a resonator made from a graphene
sheet under tensile stress at 4~K for large driving forces. The
length of the sheet is $1.7$~$\mu$m and the width is 120~nm. (a, b)
The schematics show the amplitude of motion as a function of driving
frequency (solid lines) for both sweeping directions. The dashed
lines are the solutions of the Duffing equation. (c, d) Frequency
response of the mixing current at $V^{AC}=500$~$\mu$V. The frequency
is swept upwards in (c) and downwards in (d). (Adapted with
permission from Ref.~\cite{naturenano}.)} \label{moser_fig6}
\end{figure}

For sufficiently large driving forces, the motional amplitude as a
function of the driving frequency $f$ develops an asymmetry (dashed
line in Fig.~\ref{moser_fig6}a, b). This results in bistability and
hysteresis for certain intervals in $f$ (solid line in
Fig.~\ref{moser_fig6}a, b) \cite{sazanova,chen}. The hysteresis is
intimately related to the resonance shift and also originates from
the Duffing force \cite{cross_lifshitz,kozinsky,aldridge,quirin}. An
example of a hysteretic resonance lineshape for a graphene resonator
is shown in Fig.~\ref{moser_fig6}c, d. Surprisingly, however, we do
not observe a hysteresis in some of our nanotube and graphene
resonators. This is particularly intriguing since hysteretic
behaviors are always observed in larger mechanical resonators.

The occasional absence of hysteresis is a direct consequence of
nonlinear damping. It can be predicted from the ratio between the
constants of the nonlinear forces $\alpha$ and $\eta$. When
$\eta/\alpha>\sqrt{3}/2\pi f_{0}$, the nonlinear damping is strong
enough to keep the broadening of the resonance always comparable to
or larger than its shift and precludes hysteresis for all driving
forces \cite{cross_lifshitz}. In the opposite case (when
$\eta/\alpha<\sqrt{3}/2\pi f_{0}$), a hysteretic behavior is
expected to emerge.

In the previous section we saw how $\eta$ can be extracted from the
drive dependence of the quality factor. The parameter $\alpha$ is
evaluated by fitting the shift of the resonant frequency as a
function of $V^{AC}$ to the steady-state solution of
equation~(\ref{moser_eq3}) (see solid line in
Fig.~\ref{moser_fig5}b). By comparing $\eta/\alpha$ to
$\sqrt{3}/2\pi f_{0}$, we can predict the occurrence of the
hysteresis. The agreement between experiment and theory is good in
most cases. This provides the second experimental fact that supports
the relevance the nonlinear damping force.

\section{Parametric excitation}

\begin{figure}[b]
\begin{center}
\includegraphics{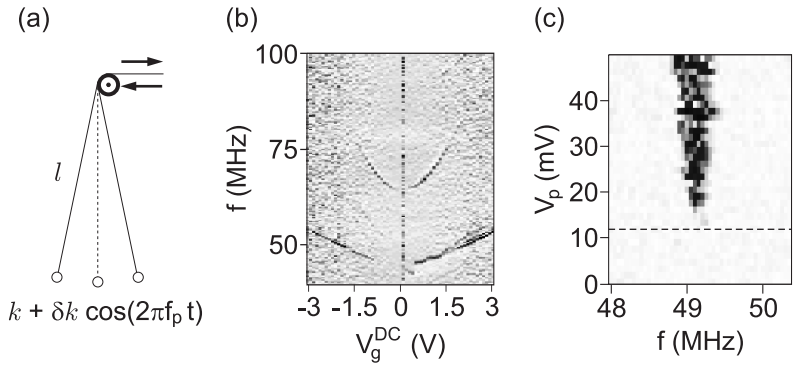}
\end{center}
\caption{(a) Schematic of a pendulum. By periodically modulating the
length $l$, one also modulates the spring constant $k+\delta
k\cos(2\pi f_{p}t)$ with $f_{p}$ an integer number of the resonant
frequency $f_{0}$ ($k=mg/l$, where $m$ is the mass of the pendulum
and $g$ is the constant of gravitation). In so doing, the pendulum
can be brought to resonance. (b) Resonant frequency as a function of
gate voltage (data obtained by measuring $I_{mix}$ versus $f$ and
$V_{g}^{DC}$). Two mechanical modes can be seen. (c) Parametric
excitation signal (dark) obtained by measuring $I_{mix}$ as a
function of the detection frequency $f$ and $V_{P}$. The phase
difference between $V_{P}$ and $V_{sd}^{AC}$ (the drive signal) is
not kept fixed so the measured $I_{mix}$ is fluctuating.
$V_{sd}^{AC}=1.4$~mV. (Adapted with permission from
Ref.~\cite{alex_nanolett}).} \label{moser_fig7}
\end{figure}

A whole new range of interesting effects becomes accessible by
modulating the parameters that enter the Newton equation. For
example, nonlinear experiments can be carried out by periodically
modulating the spring constant $k+\delta k\cos(2\pi f_{p}t)$. In the
case of a pendulum, the spring constant can be changed by varying
the length of the pendulum's arm (Fig.~\ref{moser_fig7}a). The
pendulum can then be brought to resonance by setting $f_{p}$ to be
an integer multiple of $f_{0}$, while keeping the driving force to
zero. This is what we refer to as "parametric excitation".

Nanotube resonators are expected to be excellent candidates for
parametric excitation because $k$ can be modulated with the voltage
applied to the gate by a very large amount: it is possible to make
this modulation larger than in any other mechanical resonators
fabricated to date (this can be quantified by measuring the gate
voltage dependence of the resonant frequency, which scales as
$\sqrt{k}$). Figure~\ref{moser_fig7}b shows two clearly resolved
resonant modes. Their resonant frequency can be tuned with
$V_{g}^{DC}$ to a large extent. This behavior has been attributed to
the increase of the elastic tension that builds up in the nanotube
as it bends towards the gate with increasing $V_{g}^{DC}$
(Ref.~\cite{sazanova}), in a similar fashion as a bent guitar string
vibrates at higher frequency.

In order to realize parametric excitation, we apply an oscillating
voltage $V_{P}$ at a frequency $2f$ to the gate electrode (another
excitation scheme has recently been proposed by some of us
\cite{midtvedt} and consists in applying a symmetric source-drain
voltage). On resonance, this modulates $k$ at $2f_{0}$, thereby
achieving parametric pumping of the resonator. The resulting motion
is detected using the so-called two-source technique
\cite{sazanova}, that is, by applying a small voltage $V_{sd}^{AC}$
at a slightly detuned frequency $(f-\delta f)$ to the source
electrode, and by measuring the mixing current $I_{\rm{mix}}$ at
frequency $\delta f$ at the drain electrode using a lock-in
amplifier.

Upon increasing the parametric pump excitation above a critical
threshold $V_{P,C}$, the nanotube is observed to oscillate without
any driving force (Fig.~\ref{moser_fig7}c). The resonator becomes
unstable, so that any fluctuation will activate an oscillating
motion that is sustained by the parametric drive
\cite{cross_lifshitz}. Figure~\ref{moser_fig7}c shows mechanical
motion in a tongue-shaped region of the $(f,V_{P})$ space, which is
a typical signature of self-oscillation
\cite{cross_lifshitz,turner,shim,mahboob,suh,karabalin}.
Self-oscillation is observed for $V_{P}$ roughly above
$V_{P,C}=10$~mV. In \cite{alex_nanolett} we describe additional
measurements (parametric amplification measurements) of the same
device, which independently yields $V_{P,C}\simeq10$~mV.

One interesting outcome of this experiment is that the estimation of
$V_{P,C}$ allows to extract the constant $\gamma$ of the linear
friction force $-\gamma dx/dt$ and the associated quality factor
$Q_{0}=2\pi f_{0}m/\gamma$ since below threshold the motional
amplitude is so small that nonlinear damping can be neglected. Here,
we use the simple relation
\begin{equation}
Q_{0}=\frac{1}{V_{P,C}}\cdot\frac{f_{0}}{df_{0}/dV_{g}^{DC}}
\label{moser_eq6}
\end{equation}
where $df_{0}/dV_{g}^{DC}$ is the slope of the $V_{g}^{DC}$
dependence of $f_{0}$, and obtain a quality factor $Q_{0}\simeq
1000$. This is significantly larger than the quality factor obtained
when the parametric pump excitation is off (where the quality factor
is extracted from the resonance lineshape of the resonator driven by
$F_{\rm{drive}}\cos(2\pi ft)$). There, the quality factor is about
$170-350$ depending on the driving force (as in
section~\ref{moser_section3}). We attribute this difference to the
distinct damping forces at work: $-\gamma dx/dt$ and $-\eta
x^{2}dx/dt$. While $V_{P,C}$ extracted from parametric excitation
measurements is a measure of $\gamma$, the main damping channel is
associated to the force $-\eta x^{2}dx/dt$. This is an additional
experimental fact in favor of the nonlinear damping scenario in
nanotube and graphene resonators.

\section{Discussion on the nonlinear damping force}

We have shown that the dynamics of graphene and carbon nanotube
mechanical resonators is highly unusual: the quality factor depends
on the amplitude of the motion. We emphasize however that the
quality factor can in principle saturate at low driving force for
the two following reasons. First, the quality factor is expected to
be independent of the driving force when the amplitude of thermal
vibrations is much larger than that of the driven motion, even if
the damping is dominated by the $-\eta x^{2}dx/dt$ force (M. Dykman,
private communication). Owing to the small mass of nanotubes and
graphene sheets, the amplitude $x_{th}$ of the thermal vibrations is
large. Using
\begin{equation}
x_{th}=\frac{1}{2\pi f_{0}}\sqrt{\frac{k_{\rm{B}}T}{m}}
\label{moser_eq7}
\end{equation}
yields $x_{th}=1.4$~nm at 300~K for a nanotube with $f_{0}=100$~MHz
and $m=5$~ag (which is the mass of a nanotube $1$~$\mu$m in length
and $2.2$~nm in diameter). The amplitude of the vibrations of driven
nanotube resonators was estimated from the measurements to be
typically $1-10$~nm at room temperature
\cite{sazanova,witkamp,alex_nanolett}. As such, the quality factor
is expected to rapidly saturate at 300~K at low driving forces (even
if the linear therm $-\gamma dx/dt$ is negligible).

Naturally, a second mechanism that leads to the saturation of the
quality factor at low drive is the crossover from a nonlinear to a
linear damping regime.  Provided that $\Delta f$ and the resonance
shift are much smaller than $f_{0}$, the standard definition of $Q$
is still warranted and reads $Q=2\pi E/\Delta E$ where $\Delta E$ is
the mechanical energy dissipated over one oscillation period and $E$
is the corresponding stored energy when the resonator is driven
resonantly. We obtain
\begin{equation}
1/Q=\frac{\gamma}{2\pi
f_{0}m}\left(1+\frac{\eta}{4\gamma}x_{0}^{2}\right)
\label{moser_eq8}
\end{equation}
which shows that the quality factor depends on the driving force at
large $x_{0}$ whereas it becomes a constant at low $x_{0}$ (here,
$x_{0}$ is the maximum steady state amplitude on resonance).

\begin{figure}[t]
\begin{center}
\includegraphics{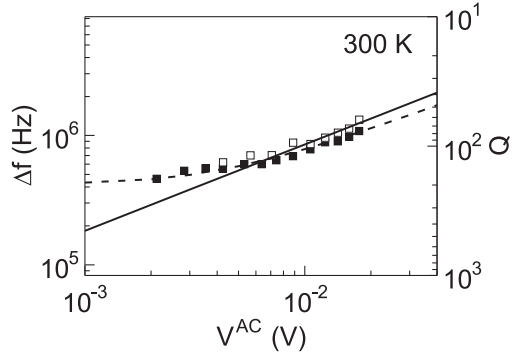}
\end{center}
\caption{Measurements of the mechanical properties of a resonator
made from a nanotube at 300~K. Resonance width and quality factor as
a function of driving force measured with the FM technique (filled
squares) and 2-source technique (hollow squares), respectively. The
solid line represents a comparison to equation~(\ref{moser_eq5})
with negligible linear damping
($\eta=2.5\cdot10^{3}$~kg$\cdot\rm{m}^{-2}\cdot\rm{s}^{-1}$,
$\gamma=0$), the dashed line is obtained with a finite $\gamma$
($\eta=1\cdot10^{3}$~kg$\cdot\rm{m}^{-2}\cdot\rm{s}^{-1}$,
$\gamma=1.9\cdot10^{-14}$~kg$\cdot\rm{s}^{-1}$). (Adapted with
permission from Ref.~\cite{naturenano}.)} \label{moser_fig8}
\end{figure}

Figure~\ref{moser_fig8} shows that the quality factor of a nanotube
resonator measured at 300~K saturates at low driving force. The
dashed line is a fit to the model that assumes the occurrence of a
crossover from the nonlinear to the linear damping regime using
$\gamma$ and $\eta$ as fit parameters (we note that the device is
different from that in Fig.~\ref{moser_fig5}). We emphasize that the
data can alternatively be captured by a model that assumes that the
saturation is due to the thermal vibrations. More work is required
to identify the proper contributions of each of these two effects.

\begin{figure}[t]
\begin{center}
\includegraphics{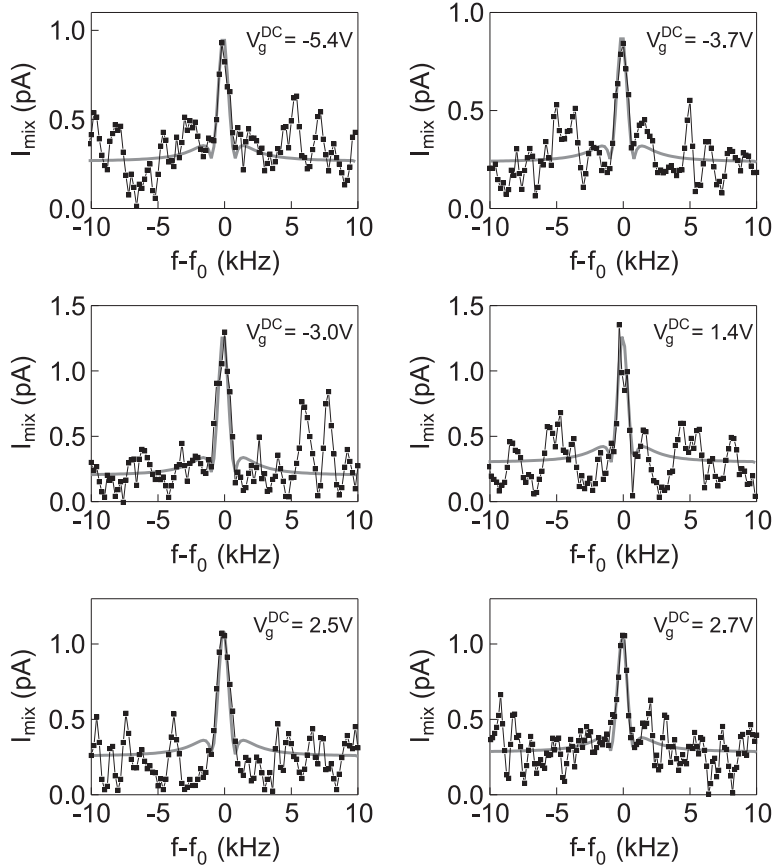}
\end{center}
\caption{Frequency response of the mixing current for a graphene
resonator. Each panel corresponds to a given gate voltage. The solid
lines are fits with $Q=10^{5}$. Measurements are carried out at
90~mK with $V^{AC}=8$~$\mu$V. The integration time of the lock-in
amplifier is 300~ms. $f_{\Delta}=1$~kHz and similar results are
obtained with $f_{\Delta}=500$~Hz. The resonance width is typically
1.5~kHz (and occasionally 300~Hz smaller or larger). (Adapted with
permission from Ref.~\cite{naturenano}.)} \label{moser_fig9}
\end{figure}

The theoretical foundations of damping rest on Newton's Principia.
Damping has been successfully described by the linear damping force
$-\gamma dx/dt$ for all mechanical resonators studied so far in
vacuum. Remarkably, this picture holds for resonators whose
dimensions span many orders of magnitude down to a few tens of
nanometers. Reducing dimensions to the atomic scale using graphene
and nanotube resonators, our work demonstrates that the simple
linear damping scenario ceases to be valid and damping is often
better described by incorporating $-\eta x^{2}dx/dt$.

This finding has profound consequences in light of the fact that
several predictions regarding quantum and sensing experiments assume
that damping is linear. Such experiments, if carried out with
nanotube or graphene resonators, would indeed have to be interpreted
within the framework of nonlinear damping. These include studies of
the quantum-to-classical transition \cite{katz}, the cooling
efficiency \cite{kippenberg}, the mass resolution \cite{ekinci}, and
the force sensitivity \cite{cleland_jap}.

Our results not only provide new insight into the dynamics of
nanotube and graphene resonators, they also hold promise for
technological applications. Our control over the resonance width
allows us to improve the mechanical quality factor. In order to
achieve larger $Q$-factors, we simply lower the driving force until
the motion becomes barely detectable. For this, it is convenient to
select the value of $V_{g}^{DC}$  for which the detection signal is
largest. In so doing, we measured a quality factor of $10^{5}$ for a
graphene resonator at 90~mK (see Fig.~\ref{moser_fig9}), which is
the largest $Q$ ever reported in a graphene resonator
\cite{naturenano}. Larger quality factors enable better force
sensing. We obtained a force sensitivity of
$2.5$~aN$/\sqrt{\rm{Hz}}$ using a nanotube resonator operating at
100~mK. This is within a factor of five of the best sensitivities
reported using microfabricated resonators operating at their
ultimate limit set by thermal vibrations \cite{mamin,teufel2009}.

The microscopic origin of the nonlinear damping is still to be
determined, but it could be related to such diverse phenomena as
phonon tunneling into the supports \cite{ignacio,prunnila}, sliding
at the contacts, nonlinearities involving phonon-phonon
interactions, or contamination in combination with geometrical
nonlinearities \cite{naturenano}. It would be of considerable
interest to experimentally study the dependence of the nonlinear
damping force on contamination, the clamping configuration, and the
suspended length. Theoretical work on the microscopic nature of
nonlinear damping should also prove valuable \cite{dykman,stav}.

\section{Coupling mechanical vibrations to electron transport in a nanotube resonator}

Lastly, we show that strong electromechanical nonlinearities arise
when a nanotube resonator operated in the Coulomb blockade regime
vibrates in concert with single electron hopping.

\begin{figure}[t]
\begin{center}
\includegraphics{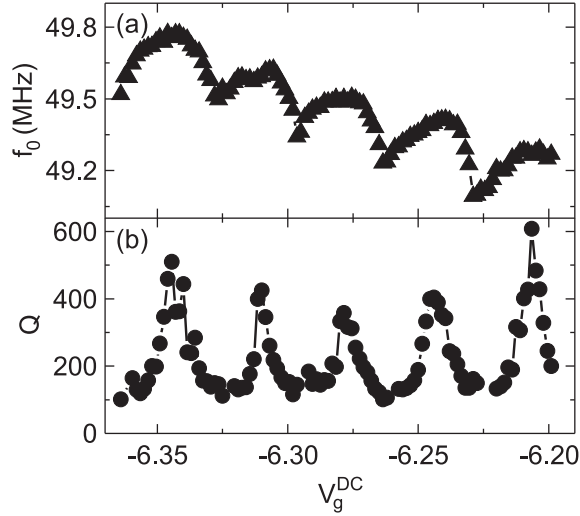}
\end{center}
\caption{(a) Resonant frequency and (b) quality factor as a function
of gate voltage $V_{g}^{DC}$ for a nanotube resonator at 4~K. The
nanotube length is $1$~$\mu$m and the diameter is $\sim 1.1$~nm.
(Adapted with permission from Ref.~\cite{benjamin_science}.)}
\label{moser_fig10}
\end{figure}

Figure~\ref{moser_fig10} shows the mechanical properties of a
nanotube resonator at 4~K. Both the resonant frequency and the
quality factor oscillate as a function of the voltage applied to the
backgate electrode. These measurements show that the dynamics of
nanotubes can be widely tuned by external electric means
($V_{g}^{DC}$), which is very advantageous for applications.

To understand the origin of the oscillations of the resonant
frequency and of the quality factor, it is useful to first consider
the electrical properties of the nanotube. The conductance
oscillates with $V_{g}^{DC}$ in a way that is typical of the
Coulomb-blockade regime and with the same $V_{g}^{DC}$ period as the
oscillations of $f_{0}$ and $Q$ (see Ref.~\cite{benjamin_science}).
This correlation indicates that the mechanical motion is affected by
Coulomb blockade.

\begin{figure}[t]
\begin{center}
\includegraphics{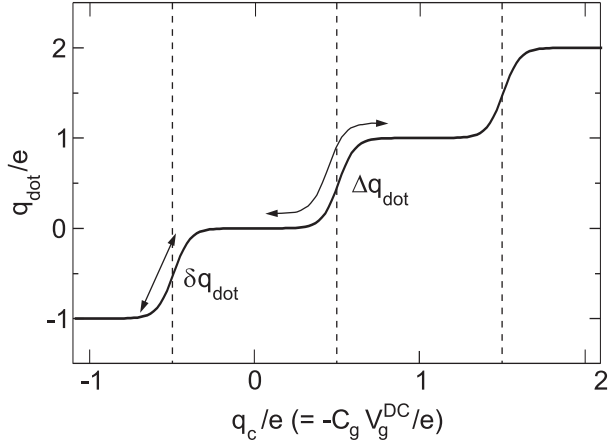}
\end{center}
\caption{Schematic of the charge on the nanotube dot as a function
of the control charge ($q_{c}=-C_{g}V_{g}^{DC}$) in the
Coulomb-blockade regime. Vibrations cause the charge on the nanotube
to oscillate (with amplitude $\delta q_{\rm{dot}}$ and $\Delta
q_{\rm{dot}}$ for low and large motional amplitude, respectively).
(Adapted with permission from Ref.~\cite{benjamin_science}.)}
\label{moser_fig11}
\end{figure}

Charge transport through nanotubes in the Coulomb-blockade regime is
a well-studied phenomenon. Upon sweeping $V_{g}^{DC}$, the
conductance oscillates as the charge $q_{\rm{dot}}$ residing on the
nanotube increases stepwise (Fig.~\ref{moser_fig11}). Furthermore,
in our experiment the nanotube behaves as a mechanical resonator.
The capacitance $C_{g}$ between the nanotube and the gate oscillates
in time as $\delta C_{g}=C_{g}^{\prime}x$ (with $C_{g}^{\prime}$ the
derivative of $C_{g}$ with respect to displacement $x$). As a
result, the nanotube is periodically charging and discharging by the
amount $\delta q_{\rm{dot}}=\delta C_{g}V_{g}^{DC}$ (as shown in
Fig.~\ref{moser_fig11}).

The oscillating charge on the nanotube results in a shift of the
resonant frequency of the resonator. When electrons tunnel onto and
out of the nanotube, the relevant electrostatic force acting on the
nanotube can be expressed as \cite{benjamin_science}
\begin{equation}
F_{e}=-\frac{C_{g}^{\prime}V_{g}^{DC}}{C_{\rm{dot}}}\left(q_{\rm{dot}}-q_{\rm{c}}\right)
\label{moser_eq9}
\end{equation}
where $q_{\rm{c}}=-C_{g}V_{g}^{DC}$, called control charge, is
introduced for commodity (this force comes from
$F_{e}=0.5C_{g}^{\prime}(V_{{\rm{dot}}}-V_{g}^{DC})^2$ with the
electrostatic potential of the nanotube dot
$V_{{\rm{dot}}}=(q_{\rm{dot}}-q_{\rm{c}})/C_{\rm{dot}}$). Because of
the repeated charging and discharging of the nanotube by the amount
$\delta q_{\rm{dot}}$, equation~(\ref{moser_eq9}) results in a
spring force, $F_{e}=-\delta k\cdot x$. This force modifies the
spring constant of the nanotube resonator by $\delta k$. As a
result, the resonator softens and the resonant frequency gets
reduced by the amount $\delta f_{0}=(f_{0}/2)\cdot(\delta k/k)$.

Besides, charging and discharging the nanotube causes damping of the
mechanical motion. Indeed, the charge $\delta q_{\rm{dot}}$ has to
flow across the tunnel barrier at the nanotube-electrode interface,
and the dissipated energy of the charge (after the tunnel process)
is supplied from the mechanical resonator.

This damping and the resonant frequency are expected to oscillate
with $V_{g}^{DC}$, since the amplitude of the $\delta q_{\rm{dot}}$
oscillations depends on $V_{g}^{DC}$. The amplitude is large (low)
when $q_{\rm{dot}}$ increases sharply (weakly) with $V_{g}^{DC}$;
see Fig.~\ref{moser_fig11}. Our model captures the data well as
discussed in \cite{benjamin_science}. We note that the background
signal of the resonant frequency in Fig.~\ref{moser_fig10}a
decreases continuously over the full $V_{g}^{DC}$ sweep: this is
attributed to the increase of the elastic tension that builds up in
the nanotube as it bends towards the gate upon increasing
$V_{g}^{DC}$ \cite{sazanova}. Second, the (oscillating) dips of the
resonant frequency in Fig.~\ref{moser_fig10}a are not symmetric,
which may be caused by the stepwise change of the electrostatic
potential when sweeping $V_{g}^{DC}$ (Ref.~\cite{steele}).

As we have seen, the coupling between mechanical vibrations and
charge transport results in an electrostatic force acting on the
nanotube \cite{benjamin_science}
\begin{equation}
F_{\rm{electro}}=-k_{\rm{electro}}x-\gamma_{\rm{electro}}dx/dt
\label{moser_eq10}
\end{equation}
where the constants $k_{\rm{electro}}$ and $\gamma_{\rm{electro}}$
can be tuned with $V_{g}^{DC}$, which is very practical for future
use. $\gamma_{\rm{electro}}$ is so large that the associated damping
force dominates over the $-\eta x^{2}dx/dt$ force.
Equation~(\ref{moser_eq9}) is valid for low $\delta q_{\rm{dot}}$.
The situation changes for large mechanical oscillation amplitude.
The charging and discharging of the dot becomes highly non-trivial
during one oscillation cycle (see $\Delta q_{\rm{dot}}$ in
Fig.~\ref{moser_fig11}). As a result, $F_{\rm{electro}}$ is expected
to depend nonlinearly on $x$ and $dx/dt$ (in contrast to the linear
dependences at low oscillation amplitude) and $F_{\rm{electro}}$ has
to be solved numerically \cite{benjamin_science}.

\begin{figure}[t]
\begin{center}
\includegraphics{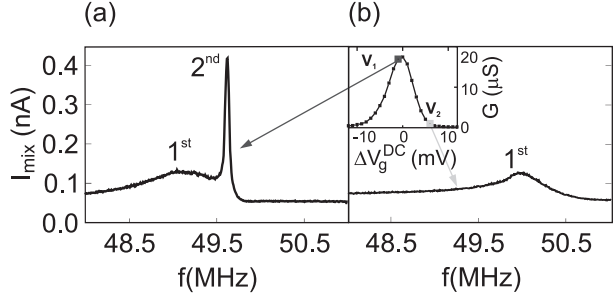}
\end{center}
\caption{Mixing current $I_{mix}$ as a function of driving frequency
for two different gate voltages at 1.5~K. A second, narrow peak
appears in (a). The inset shows the nanotube conductance as a
function of gate voltage $V_{g}^{DC}$. $\Delta V_{g}^{DC}$ is
measured from the maximum of the Coulomb-blockade peak. The nanotube
is the same as in Fig.~\ref{moser_fig10}. (Adapted with permission
from Ref.~\cite{benjamin_science}.)} \label{moser_fig12}
\end{figure}

We found evidences for nonlinearities in $F_{\rm{electro}}$.
Figures~\ref{moser_fig12}a,b show two resonance peaks at 1.5~K
(mixing current as a function of driving frequency measured using
the two-source technique \cite{sazanova}). The resonance in
Fig.~\ref{moser_fig12}a has a surprising lineshape --it splits into
two peaks, one broad and the other narrow. The narrow peak appears
only at gate voltages which yield a Coulomb blockade peak in the
conductance (inset of Fig.~\ref{moser_fig12}b). Strikingly, its
width narrows down as the driving force increases. We show in
\cite{benjamin_science} that the double-peak structure arises from
the interplay between two nonlinear phenomena, one associated with
the nonlinear force $F_{\rm{electro}}$ and the other with the
nonlinear detection mechanism. Indeed, the nonlinearity in the
detection mechanism due to Coulomb blockade can affect the lineshape
of the frequency-dependent amplitude $x(f)$ in such a way that the
lineshape of $I_{mix}(f)$ develops a double peak structure (the
reason for this is that for sufficiently large motional amplitude
$I_{mix}$ decreases when $x$ increases) \cite{benjamin_science}. If
the $x(f)$ lineshape were a Lorentzian, the double-peak structure in
$I_{mix}$ would be symmetric. However, the nonlinear force
$F_{\rm{electro}}$ renders the $x(f)$ lineshape strongly asymmetric,
which results in two peaks in $I_{mix}(f)$, one broad and the other
narrow \cite{benjamin_science}.

Overall, these results show that the coupling between mechanical
motion and electron transport in nanotube resonators can be made so
strong that the associated force acting on the nanotube becomes
highly nonlinear in its displacement and velocity. This strong
coupling originates from the reduced dimension of nanotubes. Larger
and heavier mechanical resonators are evidently much less sensitive
to the motion of individual tunneling electrons
\cite{benjamin_science}.

We note that this electro-mechanical coupling also influences charge
transport in the Coulomb blockade regime, a topic that has attracted
much interest \cite{b1,b2,b3,b4,b5,b6,b7,b8,b9,b10,b11,b12,b13,b14}.
Another active field of research consists in harnessing this
coupling to cool the mechanical motion of resonators to the phononic
ground state \cite{c1,c2,c3,c5,c6,c7}, as well as controling the
dynamics of large cantilevers of scanning probe microscopes
\cite{d1,d2,d3,d4,d5,d6}.

\section{Conclusion}

Graphene and carbon nanotubes are very interesting systems for
studying resonant mechanical behavior. They constitute the ultimate
size limit of one and two-dimensional NEMSs: nanotubes are wires
with a diameter of about 1~nm and graphene is a membrane that is
only one atom thick. Because of their reduced dimensionality,
graphene and carbon nanotubes display unusual mechanical phenomena,
among which strong nonlinearities are ubiquitous. In this chapter,
we reviewed several types of nonlinear behavior. We first discussed
an unprecedented scenario where damping is described by a nonlinear
force. This scenario is supported by several experimental facts: (i)
the quality factor varies with the amplitude of the motion as a
power law whose exponent coincides with the value predicted by the
nonlinear damping model, (ii) hysteretic behavior (of the motional
amplitude as a function of driving frequency) is absent in some of
our resonators even for large driving forces, as expected when
nonlinear damping forces are large, and (iii) when we quantify the
linear damping force (by performing parametric excitation
measurements) we find that it is significantly smaller than the
nonlinear damping force. We illustrated this chapter with
measurements on nanotube resonators but we also observed the
experimental facts (i) and (ii) on graphene resonators (see
Ref.~\cite{naturenano}). We then reviewed parametric excitation
measurements on nanotube resonators, an alternative actuation method
which is based on nonlinear dynamics. Finally, we discussed
experiments where the mechanical motion is coupled to electron
transport through a nanotube. The coupling can be made so strong
that the associated force acting on the nanotube becomes highly
nonlinear with displacement and velocity. Overall, graphene and
nanotube resonators hold promise for future studies on classical and
quantum nonlinear dynamics.

\section{Acknowledgements}
We thank M. Dykman for a critical reading of our manuscript.

\bibliography{chapter_bib}

\end{document}